\documentclass[11pt,a4paper]{article}
\usepackage{amsmath}
\usepackage{tikz}
\usetikzlibrary{calc,decorations.pathmorphing}
\usepackage{amssymb}
\usepackage{epsfig}
\usepackage{amscd}
\usepackage{graphicx,xspace}
\usepackage{float}
\usepackage[normalem]{ulem}
\usepackage{color}
\usepackage{doi}
\usepackage[T1]{fontenc}
\usepackage{authblk}

\newcommand{\twotwo}[4]{\left(\begin{array}{cc}#1&#2\\&\\#3&#4\end{array}\right)}

\title{Modified Szeg\"o-Widom Asymptotics for Block Toeplitz Matrices with Zero Modes}

\author[1]{E. Basor}

\author[2]{J. Dubail}

\author[3,4]{T. Emig}

\author[3]{R. Santachiara}

\affil[1]{American Institute of Mathematics, 600 East Brokaw Road, \protect\\
 San Jose, CA 95112, USA}

\affil[2]{Laboratoire de Physique et Chimie Th\'eoriques, \protect CNRS UMR 7019, Universit\'e de Lorraine, 
 54500 Vandoeuvre-les-Nancy, France}

\affil[3]{Laboratoire de Physique
Th\'eorique et Mod\`eles Statistiques,\protect\\ CNRS UMR 8626,
Universit\'e Paris-Saclay, Universit\'e Paris-Sud, \protect\\  91405 Orsay cedex, France}

\affil[4]{Joint MIT-CNRS Laboratory UMI 3466, \protect\\
Massachusetts Institute of Technology,  \protect\\ Cambridge, Massachusetts 02139, USA}

\begin{document}

 \maketitle

\begin{abstract}
The Szeg\"o-Widom theorem provides an expression for the determinant of block Toeplitz matrices in the asymptotic limit of large matrix dimension $n$.
We show that the presence of zero modes, i.e, eigenvalues that vanish as $\alpha^n$, $|\alpha|<1$, when $n\to \infty$, require a modification of the Szeg\"o-Widom theorem. A new asymptotic expression for the determinant of a certain class of block Toeplitz matrices with one pair of zero modes is derived. {The result is inspired by 1-dimensional topological superconductors, and the relation with the latter systems is discussed.} 
\end{abstract}

\maketitle

\section{Introduction}

{The asymptotics of the determinant of large Toeplitz matrices is often given by limit theorems \cite{BS}. Yet, subtleties can sometimes prevent one from applying known theorems in particular cases. This note is motivated by a difficulty encountered by three of us in applying the Szeg\"o-Widom theorem for block Toeplitz matrices \cite{Widom1} in a statistical physics problem connected to critical Casimir forces in a confined Ising model \cite{DSE1,DSE2}.
  
Let us start by recalling the theorem. Let $\mathbb{T} = \{ z \in \mathbb{C}; \, |z|=1 \}$ be the unit circle, and $\phi : \mathbb{T} \rightarrow \mathbb{C}^{N\times N}$ an $N \times N$ matrix valued function, called the {\it symbol}. Usually, one asks that the symbol $\phi$ satisfies certain technical assumptions regarding its regularity, but for the purposes of this note it is sufficient to work with smooth symbols, such that those assumptions hold automatically. The finite block-Toeplitz matrix $T_n(\phi)$ is defined as the matrix of total size $N n\times N n$, with an $N\times N$ block at position $i$, $j$ ($1\leq i,j \leq n$), that is the $(i-j)^{\rm th}$ Fourier coefficient of the symbol,
\begin{equation}
\phi_{i-j} = \frac{1}{2\pi i} \oint_{|z|=1} z^{i-j-1} \phi(z) dz \, .
\end{equation}
Similarly, one defines the semi-infinite block-Toeplitz matrix $T(\phi)$ with $i,j$ entry $\phi_{i-j}$ for $ 0 \leq i,j < \infty$. Then we have the

\paragraph{Szeg\"o-Widom theorem \cite{Widom1}:} {\it Let $\phi$ be a matrix-valued symbol (assumed to be smooth in this note) such that $\det \phi$ does not vanish and has winding number zero. Then
$$
\det T_n (\phi)  \, \sim \, G(\phi)^n E(\phi) , \qquad {\rm as} \; n \rightarrow \infty,
$$
where}
\begin{equation}
\label{eq:GofPhi}
G(\phi)  =  \exp \left(  \frac{1}{2\pi} \int_0^{2\pi} \log \det \phi (e^{i \theta}) d \theta  \right)
\end{equation}
{\it and $E(\phi)$ is a constant that can be expressed as $E(\phi)  =  \det T(\phi) T(\phi^{-1})$ where $T(\phi)$ and $T(\phi^{-1})$ are semi-infinite Toeplitz matrices. 
}
\vspace{0.5cm}

The theorem has a long history whose development was often motivated by a problem in mathematical physics. The first version, in the scalar case (i.e. $N=1$), was based on a question of Onsager that arose from for the 2-dimensional Ising Model \cite{McCoyWu}. It was proved by Szeg\"o \cite{Sz} in the case of positive symbols. He showed that the constant $E(\phi)$ was $$\exp \sum_{k=1}^{\infty} k s_{k}s_{-k} $$ where $s_{k}$ is the $k^{\rm th}$ Fourier coefficient of $\log \phi.$ Notice {that} in this scalar case the constant can never be zero.  Later, the theorem was generalized to more general scalar symbols, and then extended by Widom to the block case. For scalar symbols, the theorem was generalized (with a different asymptotic formula) for symbols that had jump discontinuities or vanished on the circle. For results of this type, please see \cite{Bot} where much of the history of the subject can be found.

It should also be noted that an exact identity, even in the block case, exists for the determinants for smooth symbols under certain conditions. This expression, called the Borodin-Okounkov-Geronimo-Case identity, was discovered after the original theorem was proved, but does imply the asymptotic expression. The identity is not relevant for the purposes of this paper, but for the interested reader, we refer to \cite{Bot, BW}.  \vspace{0.4cm}

Another way to state the result of the Szeg\"{o}-Widom theorem is
$$
\det T_n (\phi) / G(\phi)^n  \, \rightarrow \, E(\phi ) \qquad {\rm as} \; n \rightarrow \infty,
$$
which implies, when $E(\phi)$ is non-zero, 
$$
\frac{1}{n}  \log \det T_n (\phi) \, \rightarrow \, \frac{1}{2\pi} \int_0^{2\pi} \log \det \phi (e^{i \theta}) d \theta   \qquad {\rm as} \; n \rightarrow \infty . 
$$
In practice, in the block case, as opposed to the scalar case, the expression of $E(\phi)$ in terms of semi-infinite matrices does not allow an explicit calculation, and, in fact, it does not even allow one to decide whether or not $E(\phi)$ is zero. This can be problematic, because, if $E(\phi)$ is zero, then the Szeg\"o-Widom theorem loses its predictive value about the asymptotics of $\det T_n (\phi)$. \vspace{0.4cm}

This can be illustrated by the following examples: \vspace{0.3cm}

\noindent {\bf Example 1}. For $u \in \mathbb{R}$, consider the $2\times 2$ matrix-valued symbol
\begin{equation}
	\label{eq:symbol_C_1}
	\phi(e^{i \theta}) \, = \, \left( \begin{array}{cc}
		0  &  1 - u e^{-i \theta} \\
		-1 + u e^{i \theta} & 0 
	\end{array} \right)  ,
\end{equation}
and the associated finite block Toeplitz matrix of $n \times n$ total blocks of size $2\times 2$,
\begin{equation*}
	T_n(\phi) \, = \, \left( 
                  \vphantom{\begin{array}{c}1\\1\\1\\1\\1\\1\\1\\1\\1\\1\\1\\1\end{array}}    
                      \begin{array}{c|c|c|c|c|c}
		\begin{array}{cc} 0 & 1 \\ -1 & 0 \end{array} & \begin{array}{cc} 0 & 0 \\ u & 0 \end{array} & 0 &   & \dots & 0 \\ \hline
		 \begin{array}{cc} 0 & -u \\ 0 & 0 \end{array}  & \begin{array}{cc} 0 & 1 \\ -1 & 0 \end{array} & \begin{array}{cc} 0 & 0 \\ u & 0 \end{array} & 0 & \dots &  \\ \hline
		 0 & \begin{array}{cc} 0 & -u \\ 0 & 0 \end{array}  & \begin{array}{cc} 0 & 1 \\ -1 & 0 \end{array} & \begin{array}{cc} 0 & 0 \\ u & 0 \end{array} & 0 & \vdots \\ \hline
		  & 0 & \begin{array}{cc} 0 & -u \\ 0 & 0 \end{array}  & \ddots  & \ddots & \vdots \\ \hline		 
		 \vdots & \vdots & 0 & \ddots  & \ddots &  \begin{array}{cc} 0 & 0 \\ u & 0 \end{array} \\ \hline	 
		0 &  &  & \dots  & \begin{array}{cc} 0 & -u \\ 0 & 0 \end{array} & \begin{array}{cc} 0 & 1 \\ -1 & 0 \end{array}
	\end{array} 
              \right)
              \vspace{0.7cm}
\end{equation*}
which is real antisymmetric. It is easy to check (for instance, with an expansion by minors) that $\det T_n(\phi) = 1$, for any $n$. On the other hand, the outcome of the Szeg\"o-Widom theorem is the following. For $u \notin \{-1,1 \}$, $\det \phi (e^{i \theta}) = 1+u^2 - 2 u \cos(\theta) > 0$, so the determinant of the symbol does not vanish and has winding number zero, and the Szeg\"o-Widom theorem applies. Evaluating $G(\phi)$, one finds
\begin{equation*}
	G(\phi ) \, =\, \exp \left( \oint_{|z|=1} \frac{dz}{2 \pi i} \frac{1}{z} \log \left[(1 - u z)(1- u z^{-1})\right]  \right) \, = \,  {\rm max} (1, u^2) . 
\end{equation*}
Thus, for $-1<u < 1$, the Szeg\"o-Widom theorem predicts the correct asymptotic behavior of $\det T_n(\phi)$.  However, for $|u| > 1$, although the theorem applies, its outcome is essentially useless, because $E(\phi)$ is, in fact, zero. Then the asymptotic behavior of $\det T_n (\phi)$ is {\it not} obtained by evaluating $G(\phi)$. \vspace{0.5cm}

\noindent {\bf Example 2.} Let $u,v$ be positive real numbers with $u v <1$, and consider the symbol
\begin{equation}
	\label{eq:exact_symbol_D2}
	\phi(e^{i \theta}) \, = \,  \frac{1}{\cos \theta - \frac{uv + u^{-1}v^{-1}}{2}} \left( \begin{array}{cc}
		i \sin \theta  &   \frac{v+v^{-1}}{2}  - e^{-i \theta } \frac{u+u^{-1}}{2} \\
		-\frac{v+v^{-1}}{2}  + e^{i \theta } \frac{u+u^{-1}}{2} 	& i \sin \theta 
	\end{array} \right) ,
\end{equation}
that gives the real anti-symmetric block Toeplitz matrix of size $2n \times 2n$
\begin{equation*}
	 T_n(\phi) =  
	  {\tiny
	 \left( \begin{array}{cc|cc|c|cc}
		0 & u  & u v  & u^2 v & \dots & u^{n-1} v^{n-1} &   u^{n} v^{n-1}   \\ 
		-u & 0 & v  & u v & \dots &  u^{n-1} v^n &  u^{n-1} v^{n-1}  \\ \hline
		- uv & - v & 0 & u  &   \dots & u^{n-2} v^{n-2} &   u^{n-1} v^{n-2}   \\ 
		-u^2 v &  - uv &  -u & 0 &  \dots  & u^{n-2} v^{n-1} &  u^{n-2} v^{n-2}    \\ \hline
		- u^2 v^2 & - u v^2 & - u v  & -v &  \dots & u^{n-3} v^{n-3} &   u^{n-2} v^{n-3}    \\ 
		-u^3 v^2 &  - u^2v^2 &  -u^2 v & -u v &   \dots  &  u^{n-3} v^{n-2} &  u^{n-3} v^{n-3}    \\ \hline
		\vdots & \vdots    & \vdots & \vdots &   \ddots    &  \vdots & \vdots   \\ \hline
		-u^{n-1} v^{n-1} & - u^{n-1} v^n  &  -u^{n-2} v^{n-2} & - u^{n-2} v^{n-1}  &  \dots &  0 & u  \\
		-u^n v^{n-1}  &  -u^{n-1} v^{n-1}  & -u^{n-1} v^{n-2}  &  -u^{n-2} v^{n-2}   & \dots & -u & 0 
	\end{array} \right)  } 
\end{equation*}
Like in the previous example, this determinant can be evaluated exactly. For any $n\geq 1$, $\det T_n (\phi)\,= \, u^{2n} $. If we assume that $u \neq v$, such that 
\begin{equation*}
\det \phi(e^{i \theta}) \, = \,  \frac{(uv^{-1} + u^{-1}v)/2 - \cos \theta}{(uv + u^{-1}v^{-1})/2- \cos \theta }  >0 \, ,
\end{equation*}
then the Szeg\"o-Widom theorem applies. One finds
\begin{equation*}
	G(\phi) \,=\, \exp \left(  \oint_{|z|=1} \frac{dz}{2 \pi i}  \frac{1}{z} \log \left[ \frac{(z-u^{-1}v )(z-u v^{-1})}{(z-u v)(z-u^{-1}v^{-1})} \right] \right)  \, = \, {\rm max}(u^2,v^2)  .
\end{equation*}
Thus, the constant $E(\phi)$ in the Szeg\"o-Widom theorem is non-zero only if $u>v$. If $u<v$, the theorem does not predict the correct asymptotic behavior, because $E(\phi)=0$. \vspace{0.5cm}

These two examples show that, in general, it would be useful to have
\begin{enumerate}
	\item[(P1)] a criterion that would allow to determine whether or not, for a given matrix valued symbol $\phi$, $E(\phi)$ is zero,
	\item[(P2)] a modification of the Szeg\"o-Widom theorem for the cases with $E(\phi)=0$ that would give the correct asymptotics of $\det T_n(\phi)$.
\end{enumerate}
We shall refer to these two problems as (P1) and (P2) in the following.

The rest of this note is organized as follows. In the next section, we discuss {\bf Examples 1 and 2} from a mathematics point of view, and summarize the current knowledge about (P1) and (P2).
In section 3, we give a very brief introduction to 1d topological superconductors following Kitaev \cite{kitaev_wire}, and we explain the connection to real antisymmetric block Toeplitz matrices. In section 4, we provide precise answers to (P1) and (P2) for the class of matrices which have exactly one pair of zero modes (which includes the matrices of the so-called ``class D'' of topological superconductors in the physics literature \cite{kitaev_wire, kitaev_periodic, ryu2010topological, zirnbauer}). In the appendix we provide more examples of block Toeplitz matrices with $E(\phi) = 0$, all obtained from ``topologically non-trivial band structures''.

\section{Overview: What is known}

We start by reviewing partial answers to (P1) and (P2) from the mathematics perspective.

There is a partial answer to (P1) for certain symbols that is at times helpful. Here is what is true. If we assume the conditions of the Szeg\"o-Widom theorem and, in addition, assume that $\phi_k=0$ for all $k> a$ or that $\psi_{-k}=0$ for all $k> a$ {for some positive integer $a$}, then \[
  E(\phi)=G(\phi)^a \det T_a(\phi^{-1}).
\]

This result was originally due to Widom \cite{Widom1}, but also follows from the Borodin-Okounkov-Geronimo-Case identity mentioned above.  Notice that $\det T_a(\phi^{-1})$ is a fixed finite size matrix, so that if $a$ is not too large, it can be computed directly. Using this, one could show that for {\bf Example 1}, $E(\phi)$ vanishes for $|u| >1.$ This amounts to computing by hand a four-by-four determinant. However, this result is much less useful in {\bf Example 2}.   \\

While it is not readily clear, some insight into (P2) is gained by returning to the scalar case when we have non-zero winding. It takes only a bit of linear algebra to see that, because of the anti-diagonal structure of the symbol, the determinant of the block Toeplitz matrix in {\bf Example 1} is $(-1)^{n}$ times the product of two scalar Toeplitz determinants with symbols,

\[ \lambda(e^{i\theta}) = 1 - ue^{-i\theta},\,\,\,   \psi(e^{i\theta} )= -1 +ue^{i\theta}  . \]

{When $|u|>1$, }the first function has winding number $-1$ and the second $1$. Since both these Toeplitz determinants are triangular, their determinants are straightforward to compute. One is $(-1)^{n}$ and the other is one. Here is an alternative way to do this.

Again using some basic linear algebra it is easy to show that, if we define
\[ \xi(e^{i \theta})  = e^{-i \theta} \psi (e^{i \theta}) , \] 
then
\[ \det T_{n-1}(\psi) = \det T_{n}(\xi) \, (-1)^{n}\,(T_{n}(\xi)^{-1}(v))_{0} \]
where in the last term $v = (0,0,\dots,0, 1)$ and the subscript indicates the first component of the vector. In this case $T_{n}(\xi)^{-1} = T_{n}(\xi^{-1}).$ If we compute these terms asymptotically, we have 
\[ G(\xi)^{n} E(\xi) (-1)^{n} u^{-n} = u^{n} \times 1 \times (-1)^{n}\times u^{-n} = (-1)^{n}\]
as we know to be the case. Now, using the standard basis notation, another way to write the above is 
\[ G(\xi)^{n} E(\xi) (-1)^{n} \times \int _{0}^{2\pi}\frac{d \theta}{2\pi} e^{i n \theta}\psi^{-1}(e^{i\theta})  .\]
A similar computation can be done for $\det T_{n-1} (\lambda)$ to arrive at the asymptotic expression
\[ G(\zeta)^{n} E(\zeta) (-1)^{n} \times \int _{0}^{2\pi}\frac{d \theta}{2\pi} e^{-i n \theta}\lambda^{-1}(e^{i\theta}) \]
where $\zeta (e^{i\theta})=  e^{i \theta}\lambda(e^{i\theta}) $.

So the final answer for the original determinant in {\bf Example 1} is
\[ (-1)^{n}\,G(\xi)^{n} E(\xi) G(\zeta)^{n} E(\zeta) \times \int _{0}^{2\pi}\frac{d \theta}{2\pi} e^{i n \theta}\psi^{-1}(e^{i\theta})   \times \int _{0}^{2\pi}\frac{d \theta}{2\pi} e^{-i n \theta}\lambda^{-1}(e^{i\theta})  .\]

{We see that one way to look at this example} is to understand that, while the determinant of the symbol has no winding, there is some hidden winding. We also can see from this example that, in general, the inverse elements of $T_{n}(\xi)^{-1}$ can provide the extra factor to arrive at the correct asymptotics. In fact, whenever one can show that the leading term of  $T_{n}(\xi)^{-1}$ comes as above from the inverse of the symbol, the above formula holds in general.

The general result for scalar functions with non-zero winding is the following.
Suppose that $\psi = e^{-im\theta} \phi,$ where the winding number of $\phi$ is zero.
Then 
\begin{equation}
\det T_{n}(e^{-im\theta} \phi) \sim (-1)^{nm} \det(T_{n+m}(\phi)) F_{n,m}(\phi)
\label{th:winding_number}
\end{equation}
where this last term is $\det (T_{m}( e^{-in \theta} \alpha))$ where $\alpha$ is determined by the Wiener-Hopf factorization of $\phi$. For the interested reader, more about the theorem (\ref{th:winding_number}) and its applications can be found in \cite{BS}, Chapter 10.

The {Wiener-Hopf} factorization is of the form $\phi = \phi_{-}\phi_{+}$ where $\phi_{+}$ is a function whose Fourier coefficients vanish for negative index, $\phi_{-}$ is a function whose Fourier coefficients vanish for positive index, and  this is also true of their algebraic inverses. Another way to say this is that the `plus' function and its inverse are {in} the Hardy space, and that the `minus' function and its inverse are in the conjugate Hardy space. From the factors we define $\alpha = \phi_{-}\phi_{+}^{-1}$.
There is an analogous statement for functions with positive winding.

Before we had derived this formula in the case of  $m = \pm1$ when we decomposed {\bf Example 1} into two scalar determinants, but it can be used for similar cases such as Examples 1b and 3 in the Appendix. In general, the same analysis based on theorem (\ref{th:winding_number}) holds  in the block case when the matrix valued symbol can be factorized in the form $\phi = \phi_{-}\phi_{+}$\cite{BS}. \vspace{0.5cm}

We now turn to the more difficult and more interesting {\bf Example 2}. There, the factorization of the symbol helps to explain why things do not work exactly as we like. } Using the power decreasing method found in \cite{BB} or a direct matrix ansatz, and assuming that $uv <1$ and $u < v,$ one can show that the symbol $\phi$ of {\bf Example 2} can be factorized in the Wiener-Hopf form
\[ \phi = \phi_{+} D \phi_{-} \]
where $D$ is the diagonal matrix with $e^{-i\theta}$ and $e^{i\theta}$ on the diagonal, and the matrix valued functions are defined by
\begin{align}
\phi_+ & = \frac{-2vu}{1 - uve^{i\theta}} \twotwo{\frac{e^{2i\theta } - 1} {2} }{\frac{1}{2uv}}{\frac{e^{2i \theta } }{2}(\frac{1}{u }+ u ) - \frac{e^{i \theta }}{2} (\frac{1}{v} + v )}{\frac{1}{2u^{2}v}}
\nonumber \\
\phi_- & = \frac{1}{1 - uve^{-i\theta}}\twotwo{1}{\frac{1 }{u}}{0 } {(1-uve^{-i\theta})(ue^{-i\theta} - v)}
\, .\nonumber
\end{align}
The elements of $\phi_+$ have vanishing Fourier coefficients for negative index, and the elements of $\phi_-$ have vanishing Fourier coefficients for positive index, guaranteeing their analyticity inside and outside the unit disk in the complex plane of $z=e^{i \theta}$, respectively. This is also true, respectively, for their algebraic inverses.
If the powers of $z=e^{i \theta}$ on the diagonal of $D$ had the same winding, then a similar result to the one given above for scalar functions exists \cite{BS} (Chapter 10). But if they are different, as in this case, then it is difficult to say anything in general. We shall come back to {\bf Example 2} in Section 4 and show how the correct determinant can be computed in the asymptotic limit. \\

More will be said about the block determinants {in the next sections}, where a very similar result to the above is found, motivated by different reasoning that comes from physics. Indeed, in physics, {there exist many examples of block Toeplitz matrices for which $E(\phi) = 0$.} Those cases are inspired by 1d tight-binding models of topological insulators and superconductors, that have been studied in great detail in the physics literature since Kitaev \cite{kitaev_wire}. In addition to being physically relevant examples of block Toeplitz symbols $\phi$ with $E(\phi) = 0$, we believe that the connection with 1d topological band structures, and their classification \cite{kitaev_periodic, ryu2010topological,zirnbauer}, can shed some light on the two points above. Indeed, deciding whether $E(\phi)$ vanishes or not for a given symbol $\phi$ is, in physics, very similar to the question whether a 1d tight-binding Hamiltonian possesses zero modes, and there exist precise criteria for that in the physics literature. The question of how the Szeg\"o-Widom theorem should be modified in order to yield the correct asymptotics of $\det T_n (\phi)$ when $E(\phi)=0$ also possesses a nice reformulation in physics language: it translates to the question ``{\it How fast do the low-lying energies of a system of length $n$ go to zero as} $n \rightarrow \infty$?'', which has a precise answer, at least in simpler cases.

\section{A physics perspective: real anti-symmetric block-Toeplitz matrices and 1d topological superconductors}
\label{Szego_var}

In the { {\bf Examples 1 and 2} given in the introduction}, the block Toeplitz matrix $T_n(\phi)$ is real and antisymmetric. To some physicists, this may suggest a connection with translation invariant tight-binding models of superconductors and insulators in 1d \cite{kitaev_wire, kitaev_periodic}. We elaborate on this connection in this section.

Translation-invariant tight-binding models of fermions can be defined as follows. The degrees of freedom are self-adjoint fermions $\eta_{a,i}$, $a=1,\dots, N$, $j=0,\dots, n-1$ (also called {\it Majorana fermions}), that satisfy the canonical anti-commutation relations 
\begin{equation*}
	 \{ \eta_{a,i} , \eta_{b,j} \} = 2 \delta_{a,b} \delta_{i,j} ,\qquad \quad \eta_{a,i} = \eta_{a,i}^\dagger .
\end{equation*}
Here one thinks of $j$ as a site index along a 1d lattice, and of $a$ as an internal index on each site. The Hamiltonian of the model is of the form
\begin{equation}
	H \, = \, \frac{i}{4}   \sum_{i,j} \sum_{a,b} \eta_{a,i}   \, A_{iN + a, jN+b}  \, \eta_{b,j}
\end{equation}
where $A$ is an $nN \times n N$ real anti-symmetric matrix that has the following block structure. There is a smooth symbol $\phi : \mathbb{T} \rightarrow \mathbb{C}^{N\times N}$ such that $A$ is of the form
\begin{equation}
	\label{eq:choices}
	A \, =\, \left\{   \begin{array}{l} T_n (\phi) \qquad ({\rm open \; boundary \; conditions})  \\
	 C_n (\phi) \qquad  ({\rm periodic \; boundary \; conditions})  \end{array} \right.
\end{equation} 
where $T_n(\phi)$ is the finite block Toeplitz matrix with symbol $\phi$, and $C_n (\phi)$ is the finite {\it block circulant} matrix with the $N \times N$ block at entry $i,j$ given by the Fourier coefficients
$$
 \left\{ \begin{array}{lll}
	\phi_{i-j+n}  & {\rm if} &   i - j < -n/2 \\
	 \phi_{i-j}  & {\rm if} &   -n/2 \leq   i-j   < n/2 \\
	 \phi_{i-j-n}  & {\rm if} &   i - j \geq n/2 .
\end{array}  \right.
$$
The two choices of $A$ in Eq.~(\ref{eq:choices}) describe the same physical system in the bulk, but with different boundary conditions. \\

First, we focus on periodic boundary conditions. Because $C_n(\phi)$ is a circulant matrix, the problem of calculating its spectrum reduces to the computation of the spectrum of the $N \times N$ matrix $\sum_{p = -n/2}^{n/2-1} e^{-i p \theta} \phi_p$ which, when $n \rightarrow \infty$, is equivalent to the spectrum of the symbol $\phi (e^{i \theta})$. In particular, it is clear that, for the {\it block circulant} matrix, $\det (C_n(\phi)) \sim G(\phi)^n$ as $n \rightarrow \infty$. \\

{
Next, we turn to open boundary conditions (b.c.), and ask the question: How does the spectrum of that system (with open b.c.) differ from the previous one (periodic b.c.)?

If the two spectra are sufficiently close, that is if they are the same up to small corrections that vanish exponentially with $n$, then the asymptotics given by the Szeg\"o-Widom theorem should hold, with $E(\phi)$ non-zero. On the other hand, the case $E(\phi) = 0$ should correspond to a rather substantial change in the energy spectrum of the tight-binding model when one cuts it open.

This question is precisely the one that is at the heart of 1d topological band structures \cite{kitaev_wire, kitaev_periodic}. A Hamiltonian $H$ that has an energy gap for periodic boundary conditions can have one or more pairs of ``zero modes'' with open boundary conditions. We borrow the terminology ``zero modes" from the topological condensed matter literature, where it refers either to exact zero-energy modes or to  exponentially small tunneling modes (the case we are facing here).
 The ``zero modes'' are the eigenvalues of $T_n(\phi)$ that behave as $\lambda \sim \alpha^n$, $|\alpha| < 1$, when $n \rightarrow \infty$, that are absent from the spectrum of the circulant matrix $C_n (\phi)$. If the system has $p$ zero modes $\lambda_1  \sim \alpha_1^n$, \dots, $\lambda_p \sim \alpha_p^n $, then the Toeplitz determinant $\det T_n(\phi)$ cannot have the same asymptotics as the circulant determinant $\det C_n(\phi)$.} Instead, its behavior will be of the form
$$
\det  T_n(\phi)  \, \sim  \, \alpha_1^n \, \dots \alpha_p^n \, \det C_n(\phi) \; \times \; {\rm const.} \qquad \quad {\rm as} \quad n \rightarrow \infty .
$$
Notice that this is automatically a case where $E(\phi) = 0$ is zero in the Szeg\"o-Widom theorem.  \\

The connection with available results on 1d topological superconductors allows to go further, and to shed some light on (P1) and (P2) raised in the introduction. {\ In fact, Kitaev \cite{kitaev_wire} described all the necessary ingredients for a particular class of symbols called ``class D'' in the periodic classification of topological insulators and superconductors \cite{kitaev_periodic, ryu2010topological, zirnbauer}. For that particular class of symbols, he exhibited a precise criterion that determines whether a Hamiltonian has zero modes, solving (P1), and provided a method to evaluate the corresponding decay rates $\alpha_j$, solving (P2).

That particular class of symbols is the one of real anti-symmetric matrices with {\it no additional symmetries}. Roughly speaking, this corresponds to the case where the symbol satisfies
\begin{equation}
	\phi(e^{i \theta}) \, = \, \phi^*(e^{-i \theta}) \, = \, -\phi^\dagger(e^{i \theta}) ,
\end{equation}
and has {\it no additional block structure}. In particular, it must be neither block diagonal nor block anti-diagonal. Because the matrix $T_n(\phi)$ is antisymmetric, its eigenvalues come in complex-conjugated pairs. Kitaev argued that, for such symbols, there is either no zero mode, or exactly one pair of complex-conjugated zero modes. 

While both our examples in the introduction have exactly one pair of zero modes, only {\bf Example 2} belongs to this ``class D''. Indeed, the symbol in {\bf Example 1} is anti-diagonal and thus belongs to another symmetry class.} In this ``class D'', the existence/non-existence of a pair of zero modes is decided by the value of Kitaev's $\mathbb{Z}_2$ index \cite{kitaev_wire},
\begin{equation}
	\mathcal{I}_{\rm D} \, = \, {\rm sgn} [ {\rm Pf} \, \phi(1) \,   {\rm Pf} \, \phi(-1) ] ,
\end{equation}
where ${\rm Pf} \, M$ is the Pfaffian of the real anti-symmetric matrix $M$.
When $\mathcal{I}_{\rm D} = +1$, there are no zero modes. When $\mathcal{I}_{\rm D} = -1$, there is exactly one pair of zero modes with eigenvalues $\lambda = \pm i \varepsilon$ for $\varepsilon \in \mathbb{R}_{>0}$. { This is the answer to (P1) provided by Kitaev \cite{kitaev_wire} for ``class D'' symbols.

He also sketched an answer to (P2), by answering the physics question: How fast do the zero modes go to zero? Indeed, he argued that, for finite $n$, the reason why the energies $\pm \varepsilon$ are not exactly zero is the non-zero overlap between the two wavefunctions of the zero modes, which induces tunneling between them, and therefore energy splitting. This energy splitting is proportional to the tunneling rate, which itself is proportional to the overlap between the left and right zero modes. Therefore, the energy splitting must scale as $\varepsilon \sim e^{-n/\ell}$, where $\ell$ is the decay rate of the components of the wavefunction of each of the two zero mode.

In the next section we revisit this physics argument in more detail and discuss the modification of the Szeg\"o-Widom theorem that is needed in the presence of a pair of zero modes.}

\section{Modified Szeg\"o-Widom asymptotics for Toeplitz matrices with one pair of zero modes}

{To correct the asymptotics given by the Szeg\"o-Widom theorem in the presence of a pair of zero modes, we need to evaluate the decay rate $\alpha_1$. To do this, we make the following observation.} When $n \rightarrow \infty$, the matrix $T_n(\phi)^{-1}$ has one pair of eigenvalues that diverge, corresponding to the pair of zero modes, and all other eigenvalues are bounded. This allows us to use the {\it power iteration method} to construct the eigenvector that corresponds to the zero modes. For $n$ large, the eigenvalue is so large that a single application of $T_n (\phi)^{-1}$ is sufficient. {In other words,} for any vector $v_0 \in \mathbb{C}^{N n}$ that has non-zero overlap with the zero mode, the new vector
$$
v_1 \, = \,  T_n(\phi)^{-1} v_0
$$
is the eigenstate we are looking for, when $n \rightarrow \infty$. For $v_0$, one can generically chose one of the basis vectors $e_{iN+a}$ ($i = 1, \dots , n$, $a=1,\dots,N$) of $\mathbb{C}^{N n}$. While in general calculating the inverse of $T_n(\phi)$ is difficult, the crucial point is that the decay rate of the components of $T_n(\phi)^{-1} e_{i N+a}$ is generically the same as the one of the components of $C_n(\phi)^{-1} e_{i N+a}$. This is because one can imagine starting form a position $i$ somewhere in the bulk, so that it does not ``feel'' the boundary. Thus, the problem reduces to evaluating how fast the matrix element $[C_n (\phi)^{-1}]_{iN+a,  jN+b} =  e_{iN+a} C_n (\phi)^{-1} e_{jN+b}$ tends to zero as the distance $|i-j|$ increases. This matrix element is obtained from the Fourier transform of the inverse of the symbol, thus
$$
\int_0^{2\pi} \frac{d\theta}{2\pi} e^{i (i-j) \theta} [\phi^{-1} (e^{i \theta})]_{a b}  \; \sim \; e^{-(i-j) /\ell} \times {\rm const.}
$$
and the decay rate $\ell$ is expected to be generically the same for all matrix elements $(a,b)$. Thus, if there is exactly one pair of zero modes we have 
$$
\varepsilon^2 \, \propto \,  e^{- 2 n /\ell}  \, \propto \,  \det \left( \int_0^{2\pi} \frac{d\theta}{2\pi} e^{i n \theta} [\phi^{-1} (e^{i \theta})]_{a b} \right) \;  \; $$
as $n \rightarrow \infty$.

In summary, these arguments lead us to the following modification of the Szeg\"o-Widom theorem for the asymptotics of the determinant of $T_n(\phi)$ (which was used by three of us in the context of Refs.~\cite{DSE1,DSE2}): \\

{\bf Modified Szeg\"o-Widom asymptotics.} {\it For a Toeplitz matrix with exactly one pair of zero modes the Szeg\"o-Widom theorem is not sufficient to determine the asymptotics of $\det T_n(\phi)$, because $E(\phi) = 0$. The correct asymptotics is then given by
\begin{equation}
	\label{eq:Szego_as2}
	 \det T_n(\phi) \sim   G(\phi)^n   \; \det \left[ \int_0^{2\pi} \frac{d\theta}{2\pi}  e^{i n \theta}  \phi^{-1} (e^{i \theta}) \right] \,  \tilde E(\phi) \qquad {\rm as} \quad n \rightarrow \infty, 
\end{equation}
where the constant $\tilde E(\phi)$ is not zero. Here $G(\phi)$ is given by Eq.~\eqref{eq:GofPhi}.} \\

We emphasize that all $T_n(\phi)$ in ``class D'' with index $\mathcal{I}_{\rm D} = -1$ have exactly one pair of zero modes. But in general, the class of matrices with exactly one pair of zero modes is larger, {and this modification solves the problem in these cases as well. Notice that (\ref{eq:Szego_as2}) solves the problem of both examples in the introduction.

Indeed, for {\bf Example 2}, one finds that $\mathcal{I}_{\rm D} \, = \, {\rm sign } \left[ u-v\right] $, so the Kitaev $\mathbb{Z}_2$ index correctly predicts that $E(\phi)=0$ when $v>u$, and the asymptotics is correctly predicted by Eq.~(\ref{eq:Szego_as2}). The matrix elements of the Fourier transform of $\phi^{-1}(\theta)$ all decay as $(u v^{-1})^n$, because the decay rate is set by the complex zero of $\det \phi (\theta)$ that is closest to the real axis. As a consequence, $\det \left[ \int \frac{d\theta}{2\pi} e^{i  n \theta} \phi^{-1}(\theta) \right]  \, \sim \, (u v^{-1})^{2 n}$ up to a non-zero multiplicative constant. Hence, Eq.~(\ref{eq:Szego_as2}) correctly predicts $\det T_n(\phi) \sim  G(\phi)^n (u/v)^{2n}$, in agreement with the exact finite-size result $\det T_n (\phi) = u^{2n}$.

Although {\bf Example 1} is not in ``class D'' because it has an additional block-structure, it possesses exactly one pair of zero modes, and the correct asymptotics is also given by Eq.~(\ref{eq:Szego_as2}). }

\section{Conclusion}

Motivated by a problem encountered by three of us in applying the Szeg\"o-Widom theorem \cite{Widom1} to a statistical physics problem \cite{DSE1, DSE2}, we have pointed out an interesting phenomenon that is specific to the block Toeplitz case, and has no analog in the scalar case: Even when $\det \phi$ has no winding, $E(\phi)$ can vanish and can lead to modified versions of the Szeg\"o-Widom theorem. As an example for such a modification, we have given an explicit expression for the asymptotic determinant in the case of block Toeplitz matrices with one pair of zero modes. We also pointed out an interesting connection with 1d topological insulators and superconductors.

\paragraph{Acknowledgements.} We would like to thank E. Ardonne, K. Kozlowski, J.-M. St\'ephan and D. Vodola for inspiring discussions. This work was partly supported by the A*MIDEX Project ANR-11-IDEX-0001-02 cofunded by the French program Investissements d'Avenir, managed by the French National Research Agency.

\appendix

\section{More examples of block Toeplitz matrices with $E(\phi) = 0$: other symmetry classes of topological superconductors and insulators}
\label{sec:symmetry_classes}

At this point, we would like to come back to {\bf Example 1}. We note that it does not belong to symmetry class D because it has a symbol that is block anti-diagonal. In fact, in physics, there is another symmetry class for symbols like that in {\bf Example 1}, called class BDI \cite{kitaev_periodic}. This class is the one of real anti-symmetric matrices $A$ for which there exists a unitary matrix $M$ that squares to the identity, $M^2 = I_{Nn}$, such that
$$
M^\dagger A M \, = \, - A ,
$$
a property that encodes time-reversal symmetry. Such a property implies that, up to a change of basis, the symbol corresponding to $A$ can be transformed in block anti-diagonal form
\begin{equation}
	\label{eq:symbol_BDI}
	\phi(\theta ) \, = \, \left(  \begin{array}{c|c}
		0 & B(e^{i \theta}) \\ \hline
		- B^\dagger(e^{i \theta}) & 0
	\end{array}  \right) \, .
\end{equation}
with $B^* (e^{-i \theta}) = B(e^{i \theta})$. In class BDI, it is not Kitaev's $\mathbb{Z}_2$ invariant that decides whether or not there are pairs of zero modes. Instead, the number of such pairs is counted by the (absolute value of the) winding of $\det B(e^{i \theta})$, which we denote as $\mathcal{I}_{\rm BDI}$. The winding $\mathcal{I}_{\rm BDI}$ is an arbitrary integer number, and so the number of pairs of zero modes can be larger than 1. For instance, consider the  \\

\noindent {\bf Example 1b.} Let $u$ be  a real number, and
\begin{equation}
	\phi(\theta ) \, = \, \left(  \begin{array}{cc}
		0 &  1-u e^{-i 2 \theta}  \\
		  -1+u e^{i 2\theta} & 0   
		 \end{array}  \right)  .
\end{equation}
The corresponding Toeplitz matrix
\begin{equation*}
	T_n(\phi) \, = \, 
                     \left( \begin{array}{c|c|c|c|c|c}
		\begin{array}{cc} 0 & 1 \\ -1 & 0 \end{array} & 0 &  \begin{array}{cc} 0 & 0 \\ u & 0 \end{array} &  0 & \dots & 0 \\ \hline
		0  & \begin{array}{cc} 0 & 1 \\ -1 & 0 \end{array} & 0& \begin{array}{cc} 0 & 0 \\ u & 0 \end{array}  & \dots &  \\ \hline
		  \begin{array}{cc} 0 & -u \\ 0 & 0 \end{array}   &  0  & \begin{array}{cc} 0 & 1 \\ -1 & 0 \end{array} &0& \ddots & \vdots \\ \hline
		0  &  \begin{array}{cc} 0 & -u \\ 0 & 0 \end{array} & 0  & \ddots  & \ddots & \begin{array}{cc} 0 & 0 \\ u & 0 \end{array} \\ \hline		 
		 \vdots & \vdots & \ddots & \ddots  & \ddots & 0 \\ \hline	 
		0 &  & \dots  & \begin{array}{cc} 0 & -u \\ 0 & 0 \end{array}   &  0 & \begin{array}{cc} 0 & 1 \\ -1 & 0  \end{array}
		\end{array} \right)
\end{equation*}
has determinant $1$, for all $u$. For $u \notin \{ 1,-1 \}$, one has $\det \phi (\theta) \, = \, 1 + u^2 -2 u \cos 2\theta > 0$. Kitaev's $\mathbb{Z}_2$ index is equal to $\mathcal{I}_{\rm D} \, =\, +1$ but is does not provide sufficient information on the existence of zero modes. Instead, the proper index to use is $ \mathcal{I}_{\rm BDI}$, the winding of $\det B (e^{i \theta}) = 1 - u e^{-i 2 \theta}$, which equals $- 2$ if $|u| > 1$. Therefore, there are {\it two} pairs of zero modes in this example, when $|u|>1$.  \\

The factorization corresponding to {\bf Example 2} in Section 1 also illustrates why there is difficulty with the original Szeg\"o theorem. However here the special structure does allow one to find the formulas. Here is the factorization:
\[ \phi = P_{1} P_{2} P_{3} P_{4}\]
where 
\[ P_{1} = \twotwo{1}{0}{0}{1- e^{-i 2\theta}/u},  \quad  P_{2} = \twotwo{0}{u}{u}{0} ,  \quad  P_{3} = \twotwo{e^{i2\theta}}{0}{0}{e^{-i2\theta}}\]
and \[P_{4} = \twotwo{1}{0}{0}{-1 + e^{i2\theta}/u}.\]

The above examples illustrate that, in order to determine whether a block-Toeplitz matrix has zero modes, one needs to carefully identify the symmetries of the matrix. Different symmetries will lead to different topological invariants that count the zero modes, and to different asymptotics of block-Toeplitz determinants. In the three example considered so far, we dealt with real antisymmetric matrices, related to 1d superconductors. Looking at the topological classification of topological insulators and superconductors in Refs.~\cite{kitaev_periodic,ryu2010topological}, it is clear that there are other symmetry classes that host non-trivial topological phases of 1d insulators or superconductors (in 1d, those classes are: BDI, D, DIII, CII, AIII), and therefore are associated to block Toeplitz matrices with a particular structure that will correspond to cases where $E(\phi) = 0$ in the Szeg\"o-Widom theorem.

We give one last example, related to a 1d insulator in symmetry class AIII. This is the class of {\it hermitian} matrices $H$ that, up to a change of basis, are of block anti-diagonal form 
\begin{equation*}
\left(  \begin{array}{c|c}  0  & B(e^{i\theta}) \\ \hline B^\dagger (e^{i \theta}) & 0   \end{array} \right) \, .
\end{equation*}
Similarly to the class BDI, this may be viewed as a consequence of the existence of a unitary matrix $M$ that squares to the identity, $M^2 = I_{Nn}$, such that
\begin{equation}
	M^\dagger H M \, = \, - H ,
\end{equation}
a property often dubbed ``chiral symmetry'' or ``sub-lattice symmetry'' in physics. In class AIII, the number of pairs of zero modes is counted by the winding of $\det B(e^{i \theta})$, which defines an index $\mathcal{I}_{{\rm AIII}}$ that takes integer values. An example belonging to class AIII is the following. \\

\noindent {\bf Example 3.} Let $\zeta \in \mathbb{C}$, and
\begin{equation*}
	\phi(\theta ) \, = \, \left(  \begin{array}{cc}
		0 &   1+ \zeta^* e^{-i \theta}    \\
		   1+ \zeta e^{i \theta}    & 0
	 \end{array}  \right),
\end{equation*}
associated to the Toeplitz matrix
\begin{equation*}
	T_n(\phi) = \left( \begin{array}{c|c|c|c|c|c}
		\begin{array}{cc} 0 & 1 \\ 1 & 0 \end{array} & \begin{array}{cc} 0 & 0 \\ \zeta & 0 \end{array} & 0 &   & \dots & 0 \\ \hline
		 \begin{array}{cc} 0 & \zeta^* \\ 0 & 0 \end{array}  & \begin{array}{cc} 0 & 1 \\ 1 & 0 \end{array} & \begin{array}{cc} 0 & 0 \\ \zeta & 0 \end{array} & 0 & \dots &  \\ \hline
		 0 & \begin{array}{cc} 0 & \zeta^* \\ 0 & 0 \end{array}  & \begin{array}{cc} 0 & 1 \\ 1 & 0 \end{array} & \begin{array}{cc} 0 & 0 \\ \zeta & 0 \end{array} & 0 & \vdots \\ \hline
		  & 0 & \begin{array}{cc} 0 & \zeta^* \\ 0 & 0 \end{array}  & \ddots  & \ddots & \vdots \\ \hline		 
		 \vdots & \vdots & 0 & \ddots  & \ddots &  \begin{array}{cc} 0 & 0 \\ \zeta & 0 \end{array} \\ \hline	 
		0 &  &  & \dots  & \begin{array}{cc} 0 & \zeta^* \\ 0 & 0 \end{array} & \begin{array}{cc} 0 & 1 \\ 1 & 0 \end{array}
	\end{array} \right) \, .
\end{equation*}
that has determinant $1$. One has $\det \phi( \theta) > 0$ for $|\zeta | \neq 1$. It is easy to compute $G(\phi)$ of the Szeg\"o-Widom theorem; one finds $G(\phi) = {\rm max}[1, |\zeta|^2]$. Thus $E(\phi)$ is in fact zero if $|\zeta| > 1$. This is consistent with the observation that the index $\mathcal{I}_{\rm AIII}$ (the winding of $\det B(e^{i \theta})$) equals $-1$ for $|\zeta|>1$. \\

The factorization for this example is almost identical to {\bf Example 1b} except that $2 \theta $ is replaced by $\theta$. And as a final remark, the determinants of both of these examples follow from the remarks on page 6. Because of the specific structure of the matrices, both reduce to finding the asymptotics of two scalar functions with non-zero winding.

\end{document}